\documentclass{ifacconf}

\usepackage{graphicx}
\usepackage{amsmath}
\usepackage{amssymb}
\usepackage{bbm}
\usepackage{natbib}
\usepackage{har2nat}
\usepackage{algorithmic}
\usepackage{algorithm}
\usepackage{enumerate}
\usepackage{natbib}        

\def\myref#1{(\ref{#1})}

\def\bartheta{\bar{\theta}}
\def\barmu{\bar{\mu}}
\def\barlambda{\bar{\lambda}}
\def\barx{\bar{x}}

\def\ph{\text{ }}
\def\Roma#1{\uppercase\expandafter{\romannumeral#1}}

\newtheorem{thm1}{Theorem}

\newtheorem{rem1}{Remark}
\newtheorem{lem1}{Lemma}
\newtheorem{prop1}{Proposition}

\newtheorem{assum1}{Assumption}

\begin{document}
\begin{frontmatter}

\title{Distributed Random-Fixed Projected Algorithm for Constrained Optimization Over Digraphs\thanksref{footnoteinfo}}

\thanks[footnoteinfo]{This work was supported by the National Natural Science Foundation of China (61304038, 41576101), and Tsinghua University Initiative Scientific Research Program.}

\author[First]{Pei Xie},
\author[First]{Keyou You},
\author[First]{Shiji Song},
\author[First]{Cheng Wu}
\address[First]{Department of Automation, Tsinghua University, and Tsinghua National Laboratory for Information Science and Technology, Beijing, 100084, China (xie-p13@mails.tsinghua.edu.cn, \{youky,shijs,wuc\}@tsinghua.edu.cn)}

\begin{abstract}                
This paper is concerned with a constrained optimization problem over a directed graph (digraph) of nodes, in which the cost function is a sum of local objectives, and each node only knows its local objective and constraints. To collaboratively solve the optimization, most of the existing works require the interaction graph to be balanced or ``doubly-stochastic'', which is quite restrictive and not necessary as shown in this paper. We focus on an epigraph form of the original optimization to resolve the ``unbalanced'' problem, and design a novel two-step recursive algorithm with a simple structure. Under strongly connected digraphs, we prove that each node asymptotically converges to some common optimal solution. Finally, simulations are performed to illustrate the effectiveness of the proposed algorithms.
\end{abstract}

\end{frontmatter}
\section{Introduction}
Over the last decades, a paradigm shift from centralized processing to highly distributed systems has excited interest due to the increasing development in interactions between computers, microprocessors and sensors. In this work, we consider a distributed constrained optimization over a directed graph (digraph) of nodes. Without a central coordination unit, each node is unable to obtain the overall information of the optimization problem. More specifically, we focus on a problem of minimizing a sum of local objectives over general digraphs, where each node only accesses its local objective and constraints. Such problems arise in network congestion problems, where routers individually optimize their flow rates to minimize the latency along their routes in a distributed way. Other applications include non-autonomous power control, distributed estimation, cognitive networks, statistical inference, machine learning, and etc.

The problem of minimizing the sum of convex functions is extensively studied in recent years, see \citet{nedic2009distributed,lobel2011distributed,xi2016distributed}. In general, the existing distributed algorithms mainly adopt gradient (or sub-gradient) based methods to update the local estimate in each node to minimize its local objective, and a communication protocol is designed to achieve consensus among nodes. When constraints are taken into account, the distributed implementation of the well-known Alternating Directed method of Multipliers (ADMM) \citep{mota2013d} is proposed. It assumes the underlying graph to be undirected or balanced digraph, which is quite a restrictive assumption about real network topology. The same issue exists in other work such as \citet{nedic2010constrained}.

For unbalanced graphs,  \citet{nedic2015distributed} proposes an algorithm by combining the gradient descent and the push-sum consensus. The so-called push-sum protocol is primarily designed to achieve average-consensus on unbalanced graphs \citep{kempe2003gossip}. Although their algorithm can be applied to time-varying directed graphs, it only focuses on the unconstrained optimization, and the additional computational cost makes it very complicated. An intuitive method for unbalanced graphs is recently proposed by \citet{xi2016distributed}, which augments an additional ``surplus'' for each node to record the state updates. The main ideas are motivated by \citet{cai2012average}, which aims at achieving average consensus on digraphs. Unfortunately, the method in \citet{xi2016distributed} is unable to handle time-varying digraphs. Although there exists some distributed algorithms for constrained optimizations such as \citet{nedic2010constrained}, they mainly focus on abstract convex constraints and need to perform projection, which is computationally demanding if the projected set is irregular. Moreover, the existing algorithms dealing with structural constraints (not abstract) often encounter problems under unbalanced graphs.

To sum up, we notice that almost all the existing algorithms are either only applicable to unconstrained optimization or in need of balanced digraphs. To solve the two issues simultaneously, we introduce an epigraph form of the optimization problem to convert the objective function to a linear form, which can address the unbalanced case, and design a novel two-step recursive algorithm. In particular, we firstly solve an unconstrained optimization problem without the decoupled constraints of the epigraph form by using a standard distributed sub-gradient descent and obtain an intermediate state vector in each node. Secondly, the intermediate state vector of each node is moved toward the intersection of its local constraint sets by using the Polyak random algorithm \citep{nedic2011random}.  While a distributed version of the Polyak algorithm is proposed in \citet{you2016scenario}, in this paper we further introduce an additional ``projection'' toward a fixed direction to improve the transient performance. This algorithm is termed as \textit{distributed random-fixed projected} algorithm, of which the convergence is rigorously proved as well.

The rest of the paper is organized as follows. In Section \Roma{2}, we formulate the distributed constrained optimization and review the existing works. In Section \Roma{3}, we introduce the epigraph form of the original optimization to attack the unbalanced issue, and design a random-fixed projected algorithm to distributedly solve the reformulated optimization. In Section \Roma{4}, the convergence of the proposed algorithm is proved. In Section \Roma{5}, some illustrative examples are presented to show the effectiveness of the proposed algorithm. Finally, some concluding remarks are drawn in Section \Roma{6}.

\textbf{Notation}: For two vectors $a = \left[a_1,...,a_n\right]^T$ and $b = \left[b_1,...,b_n\right]^T$, the notation $a\preceq b$ means that $a_i\leq b_i$ for any $i\in\{1,...,n\}$. Similar notation is used for $\prec$, $\succeq$ and $\succ$. The symbols $1_n$ and $0_n$ denotes the vectors with all entries equal to one and zero respectively, and $e_j$ denotes a unit vector with the $j$th element equals to one. For a matrix $A$, we use $\|A\|$ and $\rho(A)$ to represent its norm and spectral radius respectively. Given a pair of real matrices of proper dimensions, $\otimes$ indicates their Kronecker product. The sub-gradient of a function $f$ with respect to an input vector $x\in\mathbb{R}^m$ is denoted by $\partial f(x)$. Finally, $f(\theta)_+=\max\{0,f(\theta)\}$ denotes the nonnegative part of $f$.

\section{Problem Formulation and Motivation}
\subsection{Distributed constrained optimization}
Consider a network of $n$ nodes to distributedly solve a constrained convex optimization
\begin{align}
\min_{x\in X}\quad&F(x)\triangleq\sum_{i=1}^n f_i(x),\notag\\
\label{equ:p0def}
\text{s.t.}\quad&g_i(x)\preceq0,\text{ }i=1,2,...,n,
\end{align}
where $ X\in\mathbb{R}^m$ is a common convex set known by all nodes, while $f_i: \mathbb{R}^m\rightarrow\mathbb{R}$ is a convex function only known by node $i$. Moreover, only node $i$ is aware of its local constraints $g_i(x)\preceq0$, where $g_i(x)=\left[g_i^{1}(x),...,g_i^{d_i}(x)\right]^T\in\mathbb{R}^{d_i}$ is a vector of convex functions.

We introduce a directed graph (digraph) $\mathcal{G} = \{\mathcal{V},\mathcal{E}\}$ to describe interactions between nodes, where $\mathcal{V}:=\{1,...,n\}$ denotes the set of nodes, and the set of interaction links is represented by $\mathcal{E}$. A directed edge $(i,j)\in\mathcal{E}$ if node $i$ can directly receive information from node $j$. We define $\mathcal{N}_i^{in}=\{j|(i,j)\in\mathcal{E}\}$ as the collection of in-neighbors of node $i$, i.e., the set of nodes directly send information to node $i$. The out-neighbors $\mathcal{N}_i^{out}=\{j|(j,i)\in\mathcal{E}\}$ are defined similarly. Note that each node is included in both its out-neighbors and in-neighbors. Node $i_k$ is said to be connected to node $i_1$ if there exists a sequence of directed edges $(i_1,i_2),...,(i_{k-1},i_k)$ with $(i_{j-1},i_j)\in\mathcal{E}$ for all $j\in\{2,...,k\}$, which is called a directed path from $i_1$ to $i_k$. A directed graph $\mathcal{G}$ is said to be \textit{strongly connected} if each node is connected to every other node via a directed path. If $A$ = $\{a_{ij}\}$ $\in\mathbb{R}^{n\times n}$ satisfies that $a_{ij}>0$ if $(i,j)\in\mathcal{E}$ and $a_{ij}=0$, otherwise, we say $A$ is a weighting matrix adapted to graph $\mathcal{G}$. Given a digraph $\mathcal{G}$ and its associated weighting matrix $A$, we say $\mathcal{G}$ is \textit{balanced} if $\sum_{j\in\mathcal{N}_i^{out}}a_{ji}=\sum_{j\in\mathcal{N}_i^{in}}a_{ij}$ for any $i\in\mathcal{V}$, and \textit{unbalanced}, otherwise.

Moreover, $A$ is \textit{row-stochastic} if $\sum_{j=1}^na_{ij}=1$ for any $i\in\mathcal{V}$, and \textit{column-stochastic}, if $\sum_{i=1}^na_{ij}=1$ for any $j\in\mathcal{V}$. The matrix $A$ is said to be \textit{doubly-stochastic} if $A$ is both row-stochastic and column-stochastic. We note that double-stochasticity is a restrictive condition for digraphs.

The objective of this paper is to design a distributedly recursive algorithm for problem \myref{equ:p0def} over an \textit{unbalanced} digraph, under which every node $i$ updates a local vector $x_i^k$ by exchanging limited information with neighbors at each time so that each $x_i^k$ eventually converges to some common optimal solution.

\subsection{Review of the major results and motivation}
\label{subsec_review}
We first review the standard distributed gradient descent algorithm (DGD) for \textit{unconstrained} optimization, which however requires \textit{doubly-stochastic} weighting matrices \citep{nedic2009distributed}. That is, each node $i$ updates its local estimate of an optimal solution by
\begin{equation}
\label{equ:dgd}
x_i^{k+1}=\sum_{j=1}^na_{ij}x_j^k-\zeta^k\nabla f_i,
\end{equation}
where $\zeta^k$ is a given step size.

However, the DGD is only able to solve the optimization problem over balanced graphs, which is not applicable to unbalanced graphs. To illustrate this point, we  define the \textit{Perron vector} of a weighting matrix $A$ as follows.

\begin{lem1}\label{lemma:perron} \citep[Perron Theorem]{horn2012matrix} If $\mathcal{G}$ a strongly-connected digraph and $A$ is the associated weighting matrix, there exists a Perron vector $\pi\in\mathbb{R}^n$ such that
\begin{equation}
\label{equ:l1res1}
\pi^TA=\pi^T, \pi^T1_n=1,\pi_i>0,~\text{and}~
\end{equation}
\begin{equation}
\label{equ:l1res2}
\rho(A-1_n\pi^T)<1.
\end{equation}
\end{lem1}
By multiplying $\pi_i$ in \myref{equ:l1res1} on both sides of \myref{equ:dgd} and summing up over $i$, we obtain that
\begin{align}
\label{equ:sdgd}
  \barx^{k+1}&\triangleq\sum_{i=1}^n\pi_ix_i^{k+1} \notag\\
             &=\sum_{j=1}^n\left(\sum_{i=1}^n\pi_i a_{ij}\right)x_j^k-\zeta^k\sum_{i=1}^n\pi_i\nabla f_i(x_i^k)\notag\\
             &=\barx^{k}-\zeta^k\sum_{i=1}^n\pi_i\nabla f_i(x_i^k).
\end{align}
If all nodes have already reached consensus, then \myref{equ:sdgd} is written as
\begin{equation}
\label{equ:cgd}
\barx^{k+1}=\barx^k-\zeta^k\sum_{i=1}^n\pi_i\nabla f_i(\barx^k).
\end{equation}
Clearly, \myref{equ:cgd} is a gradient descent algorithm to minimize the following objective function
\begin{equation}
\bar{F}(x)\triangleq\sum_{i=1}^n\pi_if_i(x).
\end{equation}
Thus, each node converges to a minimizer of $\bar{F}(x)$ rather than $F(x)$ in \myref{equ:p0def}, which is also noted in \citet{xi2016distributed}. For a generic unbalanced digraph, the weighting matrix is no longer doubly-stochastic, and the Perron vector is not equal to $\left[\frac{1}{n},...,\frac{1}{n}\right]^T$,  which obviously implies that $\bar{F}(x)\neq F(x)$. That is, DGD in \myref{equ:dgd} is not applicable to the case of unbalanced graphs.

If each node $i$ is able to access its associated element of the Perron vector $\pi_i$, it follows from \myref{equ:cgd} that a natural way to modify the DGD in \myref{equ:dgd} is given as
$$
x_i^{k+1}=\sum_{j=1}^na_{ij}x_j^k-\frac{\zeta^k}{\pi_i}\nabla f_i,
$$
which is recently exploited in \citet{moral2016Perronestimate} by designing an additional distributed algorithm to locally estimate $\pi_i$. However, it is not directly applicable to time-varying graphs as there does not exist such a {\em constant} Perron vector. In fact, this shortcoming has also been explicitly pointed out in \citet{moral2016Perronestimate}. Another idea to resolve the unbalanced problem is to augment the original row-stochastic matrix into a doubly-stochastic matrix. This novel approach is originally proposed by \citet{cai2012average} for average consensus achieving problems over unbalanced graphs. Their key is to augment an additional variable for each agent, called ``surplus", whose function is to locally record individual state updates. In \citet{xi2016distributed},  the ``surplus-based'' idea is adopted to solve the distributed optimization problem over fixed unbalanced graphs. Although it is extended to time-varying graphs in \citet{cai2014average}, it only focuses on the average consensus problem.  Again, it is unclear how to use the ``surplus-based'' idea to  solve the distributed optimization problem over time-varying unbalanced graphs. This problem has been resolved in \citet{nedic2015distributed}  by adopting the so-called push-sum consensus protocol, the goal of which is to achieve the average consensus over unbalanced graphs. Unfortunately, their algorithms appear to be over complicated and involve nonlinear iterations.  More importantly,  they are restricted to the unconstrained optimization, and their rationale is not as clear as the DGD.

In this work, we solve the unbalanced problem from a different perspective, which can easily address the constrained optimization over time-varying digraphs\footnote{The result on time-varying digraphs is to be included in the journal version of this work.}.
\section{Distributed Algorithms for Constrained Optimization}
As explained, perhaps it is not effective to attack the unbalanced problem via the Perron vector. To overcome this limitation, we study the epigraph form of the optimization \myref{equ:p0def}, and obtain the same linear objective function for every node. This eliminates the effect of different elements of the Perron vector on the limiting point of \myref{equ:cgd}. Then we utilize the DGD in \myref{equ:dgd} to resolve the epigraph form and obtain an intermediate state vector. The feasibility of the local estimate in each node is asymptotically ensured by further driving this vector toward the constraint set. That is,  we update  the intermediate vector toward the negative sub-gradient direction of a local constraint function.  This novel idea is in fact proposed in our recent work \citep{you2016scenario}, which generalizes the Polyak random algorithm to its distributed version. The convergence of the algorithm is proved in next section.
\subsection{Epigraph form of the constrained optimization}
Our main idea does not focus on $\pi_i$ but on $f_i$ in \myref{equ:cgd}. Specifically, if we transform all the local objective $f_i(x)$ to the same form $f_0(x)$, then \myref{equ:cgd} is reduced to $\barx^{k+1}=\barx^k-\zeta^k\nabla f_0(\barx^k)$, which implies that there is no difference between the cases of balanced and unbalanced digraphs. This is achieved by concentrating on the epigraph form of the optimization \myref{equ:p0def}.

Given a  $f(x):\mathbb{R}^m\rightarrow\mathbb{R}$, the epigraph of $f$ is defined as
\begin{equation*}
\textbf{epi}\ph f\ph =\ph\{(x,t)|x\in \textbf{dom}\ph f, f(x)\leq t\},
\end{equation*}
which is a subset of $\mathbb{R}^{m+1}$. It follows from \citet{boyd2004convex} that the epigraph of $f$ is a convex set if and only if $f$ is convex, and minimizing $f$ is equal to searching the minimal auxiliary variable $t$ within the epigraph. By this way, we transform the optimization problem of minimizing a convex objective to minimizing a \textit{linear function} within a convex set. In the case of multiple functions, the epigraph can be defined similarly by introducing multiple auxiliary variables.

Combining the above ideas, we consider the epigraph form of \myref{equ:p0def} by using an auxiliary vector $t\in\mathbb{R}^n$. Then, it is clear that problem \myref{equ:p0def} can be reformulated as
\begin{align}
\min_{(x,t)\in\Theta}\quad&\sum_{i=1}^n 1_n^Tt/n,\notag\\
\text{s.t.}\quad&f_i(x)-e_i^Tt\leq 0,\notag\\
\label{equ:p1def}
&g_i(x)\preceq0,\quad i=1,2,...,n,
\end{align}
where $\Theta= X\times\mathbb{R}^n$ is the Cartesian product of $ X$ and $\mathbb{R}^n$.
\begin{rem1}\label{rem:epi} In view of the epigraph form, we have the following comments.
\begin{enumerate}\renewcommand{\labelenumi}{\rm(\alph{enumi})}
\item Denote $y=[x^T,t^T]^T$ and $f_0(y)=c_0^Ty/n$, where $c_0=[0_m^T,1_n^T]^T$. Thus, the objective in \myref{equ:p1def} becomes the sum of the local objective $f_0$, which is the same for all nodes. In view of \myref{equ:cgd}, the correctness of the DGD can be guaranteed even for  unbalanced graphs.
\item The local objective $f_i(x)$ in \myref{equ:p0def} is handled via an additional constraint in \myref{equ:p1def} such that $\tilde{f}_i(y)= f_i(x)-e_i^Tt\leq0$, where $\tilde{f}_i(y)$ is a convex function as well. To evaluate $\tilde{f}_i(y)$, it requires each node $i$ to select the $i$-th element of the vector $t$ and $i$ is the identifier of node $i$.  As a result, the epigraph form requires each node to know its identifier, which is also needed in \citet[Assumption 2]{van2016distributed}.
\end{enumerate}
\end{rem1}
\subsection{Distributed Random-Fixed Projected Algorithm}
Since the objective function of \myref{equ:p1def} is linear, there does not exist an optimal point for the unconstrained optimization without local constrains in \myref{equ:p1def}. That is, the problem \myref{equ:p1def} is meaningful only when constraints are taken into consideration. Consider that the local constraints in \myref{equ:p1def} are given in the form of convex functions, we shall fully exploit their structures, which is different from the constrained version of DGD in \citet{lobel2011distributed} by using the projection operator. Clearly, the projection is easy to perform only if the projected set has a relatively simple structure, e.g., interval or half-space. From this perspective, our algorithm requires much less computational load per iteration.  To this end,  we adopt the Polyak random algorithm \citep{nedic2011random}, which however only addresses the centralized version, to solve the distributed constrained optimization.

To solve \myref{equ:p1def} recursively, every node $j$ maintains a local estimate $x_j^k\in\mathbb{R}^m$ and $t_j^k\in\mathbb{R}^n$ at each iteration $k$. Firstly, we solve an unconstrained optimization problem which removes the constraints in problem \myref{equ:p1def} by using the standard distributed sub-gradient descent algorithm and obtain intermediate state vectors $p_j^k$ and $y_j^k$, which correspond to $t_j^k\in\mathbb{R}^m$ and $x_j^k\in\mathbb{R}^n$, respectively, i.e.,
\begin{align}
\label{equ:a1p1}
p_j^k&=\sum_{i=1}^na_{ji}t_i^k-\zeta^k1_n,\\
\label{equ:a1p2}
y_j^k&=\sum_{i=1}^na_{ji}x_i^k,
\end{align}
where $\zeta^k$ is the step-size satisfying the persistently exciting condition
\begin{equation}\label{equ:stepdef}
\zeta^k>0,\quad\sum_{k=0}^\infty\zeta^k=\infty,\quad\sum_{k=0}^\infty(\zeta^k)^2<\infty.
\end{equation}
Then, we adopt the Polyak's idea to address the constraints of \myref{equ:p1def} to drive the intermediate state vectors toward the feasible set. To facilitate the presentation, we introduce following notations
\begin{align}
X_j^l&=\{x\in \mathbb{R}^m|g_j^l(x)\leq0\}, l\in\{1,...,\tau_j\},\notag\\
\label{equ:oldsetdef}
X_j^0\times T_j&=\{(x,t)|f_j(x)-e_j^Tt\leq0,x\in X\}.
\end{align}
To be specific, we update $y_j^k$ toward a randomly selected set $X_j^{\omega_j^k}$ by using the Polyak's projection idea, i.e.,
\begin{equation}
\label{equ:a1p3}
z_j^{k}=y_j^k-\beta\frac{g_j^{\omega_j^k}(y_j^k)_+}{\|u_j^k\|^2}u_j^k,
\end{equation}
where $\beta\in(0,2)$ is a constant parameter, $\omega_j^k$ is a random variable taking value from the integer set $\{1,...,\tau_j\}$, and the vector $u_j^k\in\partial g_j^{\omega_j^k}(y_j^k)_+$ if $g_j^{\omega_j^k}(y_j^k)_+>0$ and $u_j^k=u_j$ for some $u_j\neq0$ if $g_j^{\omega_j^k}(y_j^k)_+=0$. In fact, $u_j^k$ is a decreasing direction of $g_j^{\omega_j^k}(y_j^k)_+$, which leads to that
$d(z_j^k,  X_j^{\omega_j^k})\leq d(y_j^k,  X_j^{\omega_j^k})$ for sufficiently small $\beta$. If $\omega_j^k$ is appropriately selected, it is expected in the average sense that
$d(z_j^k, \cap_{l=1}^{\tau_j} X_j^l)\leq d(y_j^k, \cap_{l=1}^{\tau_j} X_j^l).$

It is noted that the auxiliary vector $t_j^k$ is not updated during the above process. We use the same idea to handle the newly introduced constraint $X_j^0\times T_j$ such that
\begin{align}
\label{equ:a1p4}
x_j^{k+1}&=\Pi_X(z_j^k-\beta\frac{(f_j(z_j^k)-e_j^Tp_j^k)_+}{1+\|v_j^k\|^2}v_j^k),\\
\label{equ:a1p5}
t_j^{k+1}&=p_j^k+\beta\frac{(f_j(z_j^k)-e_j^Tp_j^k)_+}{1+\|v_j^k\|^2}e_j,
\end{align}
where the vector $v_j^k$ is a sub-gradient of $f_j$ evaluated at $z_j^k$. Similarly, we have that
\begin{equation}
d((x_j^{k+1},t_j^{k+1}), X_j^0\times T_j)\leq d((z_j^k,t_j^k), X_j^0\times T_j).
\end{equation}
Once all the nodes reach an agreement, the state vector $(x_j^k,t_j^k)$ in each node asymptotically converges to a feasible point.  Overall, we use Algorithm \ref{alg_disrandom} to formalize the above discussion. Note that \citet{nedic2009distributed} requires the \textit{double stochasticity} of $A$, which is unnecessary here.
\begin{algorithm}[htbp!]
\caption{Distributed random-fixed projected algorithm (D-RFP)}
\begin{enumerate}[1:]
\item \textbf{Initialization:} For each node $j\in\mathcal{V}$, set $x_j=0,t_j=0$.
\item\textbf{Repeat}
\item\textbf{Set} $k=1$.
\item\textbf{Local information exchange:} Each node $j\in\mathcal{V}$ broadcasts $x_j$ and $t_j$ to its out-neighbors.
\item\textbf{Local variables update:} Each node $j\in\mathcal{V}$ receives the state vectors $x_i$ and $t_i$ from its in-neighbors $i\in\mathcal{N}_j^{in}$ and updates its local vectors as follows
\begin{itemize}
  \item $y_j=\sum_{i\in\mathcal{N}_j^{in}}a_{ji}x_i$, $p_j=\sum_{i\in\mathcal{N}_j^{in}}a_{ji}t_i-\zeta^k1_n$, where the stepsize $\zeta^k$ is given in \myref{equ:stepdef}.
  \item Draw a random variable $\omega_j$ from $\{1,...,\tau_j\}$, and obtain $z_j=y_j-\beta\frac{g_j^{\omega_j}(y_j)_+}{\|u_j\|^2}u_j$, where $u_j$ is defined in \myref{equ:a1p3}.
  \item Set $x_j\leftarrow\Pi_X(z_j-\beta\frac{(f_j(z_j)-e_j^Tp_j)_+}{1+\|v_j\|^2}v_j)$, where $v_j$ is defined in \myref{equ:a1p4}, and $t_j\leftarrow p_j+\beta\frac{(f_j(z_j)-e_j^Tp_j)_+}{1+\|v_j\|^2}e_j$.
\end{itemize}
\item\textbf{Set} $k=k+1$.
\item\textbf{Until} a predefined stopping rule is satisfied.
\end{enumerate}
\label{alg_disrandom}
\end{algorithm}
\begin{rem1}\label{rem:polyak1} Algorithm \ref{alg_disrandom}  is motivated by a centralized Polyak random algorithm \citep{nedic2011random}, which is very recently extended to the distributed version in \citet{you2016scenario}. The main difference from \citet{you2016scenario} is that we do not use randomized projection on all the constraints. For instance, $X_j^0\times T_j$ will always be considered per iteration. If we equally treat the constraints $g_j(x)\preceq 0$ and $f_j(x)-e_j^Tt\leq 0$, then once the selected constraint is from an element of $g_j(x)$, the vector $t$ is not updated as $t$ is independent of $g_j(x)$. This will slow down the convergence speed and introduce undesired transient behavior. Thus, Algorithm \ref{alg_disrandom}  adds a fixed projection to ensure that both $x$ and $t$ are updated at each iteration.
\end{rem1}
\begin{rem1}\label{rem:polyak2}We observe that Algorithm \ref{alg_disrandom} is also motivated from the alternating projection algorithm, which searches the intersection of several constraint sets by employing alternating projections, see e.g. \citet{escalante2011alternating} and references therein. The key idea of the algorithm is that the state vector will asymptotically get closer to the intersection by repeatedly projecting to differently selected constraint sets. In light of this, the ``projection''  in our algorithm can also be performed for any times at each iteration, either randomly or fixedly, to achieve the feasibility. In fact, we can also design other rules for selecting the projected constraint. For example, we may choose the most distant constraint set from the intermediate vector. The measure of the ``distance'' from a vector $x$ to a constraint set $f(x)\leq0$ is given as $\frac{f(x)_+}{\|\partial f(x)\|}$.
\end{rem1}
\section{Convergence Analysis}
To prove the convergence of Algorithm \ref{alg_disrandom}, we consider a general form of \myref{equ:p1def} to simplify notations as
\begin{align}
\min_{\theta\in\Theta}\quad&c^T\theta,\notag\\
\text{s.t.}\quad &f_j(\theta)\leq0,\notag\\
\label{equ:p2def}
&g_j(\theta)\preceq0,\quad j=1,2,...,n,
\end{align}
where $\theta=(x,t) \in\mathbb{R}^d$ in  \myref{equ:p1def} and $d=m+n$.  Moreover, $f_j:\mathbb{R}^d\rightarrow\mathbb{R}$ is a convex function and $g_j:\mathbb{R}^d\rightarrow\mathbb{R}^{\tau_j}$ is a vector of convex functions.

The objective is a global linear function, and each node maintains local constraints only known by itself. In the optimization problem \myref{equ:p2def}, the inequality $f_j(\theta)\leq 0$ is regarded as a crucial constraint that needs to be prior satisfied, while some constraints in $g_j(\theta)\preceq0$ can be temporarily relaxed until being selected. Then the D-RFP algorithm for \myref{equ:p2def} is given as
\begin{subequations}
\label{equ:generic}
\begin{align}
\label{equ:a2p1}
p_j^k&=\sum_{i=1}^na_{ji}\theta_i^k-\zeta^kc,\\
\label{equ:a2p2}
q_j^k&=p_j^k-\beta\frac{g^{\omega_j^k}(p_j^k)_+}{\|u_j^k\|^2}u_j^k,\\
\label{equ:a2p3}
\theta_j^{k+1}&=\Pi_\Theta(q_j^k-\beta\frac{f_j(p_j^k)_+}{\|v_j^k\|^2}v_j^k),
\end{align}
\end{subequations}
where $u_j^k\in\partial g_j^{\omega_j^k}(p_j^k)_+$ if $g_j^{\omega_j^k}(p_j^k)_+>0$ and $u_j^k=u_j$ for some $u_j\neq0$ if $g_j^{\omega_j^k}(p_j^k)_+=0$, and the vector $v_j^k$ is defined similarly related to $f_j$.

It is easy to verify that Algorithm \ref{alg_disrandom} is just a special case of the algorithm given in \myref{equ:generic}. Therefore, we only need to prove the convergence of  \myref{equ:generic}.  To this end,  we introduce the following notations
\begin{align}
\Theta_j&=\{\theta\in\Theta|f_j(\theta)\leq0,g_j(\theta)\preceq0\},\notag\\
\Theta_0&=\Theta_1\cap\cdots\cap\Theta_n,\notag\\
\label{def:theta}
\Theta^*&=\{\theta\in\Theta_0|c^T\theta\leq c^T\theta',\forall\theta'\in\Theta_0\}.
\end{align}
Before proceeding,  several assumptions are needed.
\begin{assum1}\label{assum:solvable} The optimization problem in  \myref{equ:p2def}  is feasible and has a nonempty set of optimal solutions, i.e., $\Theta_0\neq\varnothing$ and $\Theta^*\neq\varnothing$.
\end{assum1}
\begin{assum1}\label{assum:connec} (Strong connectivity). The graph $\mathcal{G}$ is strongly connected.\end{assum1}
Assumption \ref{assum:solvable} is trivial that ensures the solvability of the problem. As the constraints in \myref{equ:p2def} are only known to node $j$, the strong connectivity of $\mathcal{G}$ is also necessary. Otherwise, we may encounter a situation where a node $i$ can never be accessed by some other node $j$, thus the information from node $i$ cannot reach node $j$. Then, it is impossible for node $j$ to find a solution to \myref{equ:p2def} since the information on the constraints maintained by node $i$ is always missing to node $j$. To ensure the convergence of the proposed algorithm, we also need the following assumptions.
\begin{assum1}\label{assum:random} (Randomization and sub-gradient boundedness). Suppose the following holds
\begin{enumerate}[(a)]
\item $\{\omega_j^k\}$ is an i.i.d. sequence that is uniformly distributed over $\{1,...,\tau_j\}$ for any $j\in\mathcal{V}$, and is also independent over the index $j$.
\item The sub-gradient $u_j^k$ and $v_j^k$ are uniformly bounded over the set $\Theta$, i.e., there exists a scalar $D>0$ such that
    \begin{equation*}
    \max\{\|u_j^k\|,\ph\|v_j^k\|\}\leq D,\ph\forall j\in\mathcal{V},\forall k>0.
    \end{equation*}
\end{enumerate}
\end{assum1}
Obviously, the designer can freely choose any distribution for drawing the samples $\omega_j^k$. Hence Assumption \ref{assum:random}(a) is easy to satisfy. The Assumption \ref{assum:random}(b) is also common for the optimization problem, see e.g. \citet[Assumption 7]{nedic2009distributed}, which is not hard to satisfy.

Now we are ready to present the convergence result on the distributed random-fixed projected algorithm.
\begin{thm1}\label{thm:result}
(Almost sure convergence). Under Assumptions \ref{assum:solvable}-\ref{assum:random}, the sequence $\{\theta_j^k\}$ in \myref{equ:generic} almost surely converges to some common point in the set $\Theta^*$ of the optimal solutions to \myref{equ:p2def}.
\end{thm1}
The proof of Theorem \ref{thm:result} is roughly divided into three parts. The first part demonstrates the asymptotic \textit{feasibility} of the state vector $\theta_j^k$, see Lemma \ref{lemma:feasi}. The second part illustrates the \textit{optimality} by showing that the distance of $\theta_j^k$ to any optimal point $\theta^*$ is ``stochastically'' decreasing. Finally, the last part establishes a sufficient condition to ensure asymptotic \textit{consensus} in Lemma \ref{lemma:consensus}, under which the sequence $\{\theta_j^k\}$ converges to the same value for all $j\in\mathcal{V}$. By using the above results, we show that $\{\theta_j^k\}$ converges to some common point in $\Theta^*$ almost surely.
\begin{lem1}\label{lemma:toset}(Iterative projection). Let $\{h_k\}:~\mathbb{R}^m\rightarrow\mathbb{R}$ be a sequence of convex functions and $\{\Omega_k\subseteq\mathbb{R}^m\}$ be a sequence of convex closed sets. Define $\{y_k\}\subseteq \mathbb{R}^m$ by
\begin{equation*}
y_{k+1}=\Pi_{\Omega_k}(y_k-\beta\frac{h_{k}(y_k)_+}{\| d_k\|^2}d_k),
\end{equation*}
where $0<\beta<2,d_k\in\partial h_{k}(y_k)$ if $h_{k}(y_k)>0$ and $d_k= d$ for any $d\neq 0$, otherwise. For any $z\in{(\Omega_0\cap\cdots\cap\Omega_{k-1})}\bigcap\{y|h_j(y)\leq0,j=0,\ldots,k-1\}$, it holds that
\begin{equation*}
\| y_k-z\|^2\leq\| y_0-z\|^2-\beta(2-\beta)\frac{\| h_{0}(y_0)_+\|^2}{\| d_0\|^2}.
\end{equation*}
\end{lem1}
\begin{pf} By \citet[Lemma 1]{nedic2011random} and the definition of $\{y_k\}$, it holds for $j\leq k-1$ that $\| y_{j+1}-z\|^2\leq\| y_{j}-z\|^2-\beta(2-\beta)\frac{\| h_{j}(y_j)_+\|^2}{\| d_{j}\|^2}$. Together with the fact that $0<\beta<2$, we have that $\| y_{j+1}-z\|^2\leq\| y_{j}-z\|^2$. Then, $\| y_{k}-z\|^2\leq\| y_{1}-z\|^2\leq\| y_0-z\|^2-\beta(2-\beta)\frac{\| h_{0}(y_0)_+\|^2}{\| d_0\|^2}$.
\end{pf}
\begin{lem1}\label{lemma:feasi} (Feasibility). Define $\lambda_j^k$ and $\mu_j^k$ as follows
\begin{equation}
\label{equ:l4def}
\lambda_j^k=\sum_{i=1}^n a_{ji}\theta_i^k,~\text{and}~\ph\mu_j^k=\Pi_{\Theta_0}(\lambda_j^k),
\end{equation}
where $\Theta_0$ is defined in \myref{def:theta}. If $\lim_{k\rightarrow\infty}\| \lambda_j^k-\mu_j^k\|=0$ for any $j\in\mathcal{V}$, then $\lim_{k\rightarrow\infty}\| \mu_j^k-\theta_j^{k+1}\|=0$.
\end{lem1}
\begin{pf}Consider Lemma \ref{lemma:toset}, let $y_0=p_j^k$, where $p_j^k$ is given in \myref{equ:a2p1}, $h_0(y)=g_j^{\omega_j^k}(y)$ and $h_1(y)=f_j(y)$, $\Omega_0=\mathbb{R}^m$ and $\Omega_1=\Theta$. Then it is clear that $y_2=\theta_j^{k+1}$. Since $\mu_j^k\in\Theta_0\subseteq(\Omega_0\cap\Omega_1)$, both $y_0(\mu_j^k)\leq0$ and $y_1(\mu_j^k)\leq0$ are satisfied. By Lemma \ref{lemma:toset}, it holds that
\begin{equation*}
\|\theta_j^{k+1}-\mu_j^k\|^2\leq\| p_j^k-\mu_j^k\|^2-\beta(2-\beta)\frac{g_j^{\omega_j^k}(p_j^k)_+^2}{\| d_j^k\|^2}.
\end{equation*}
Notice that $\| p_j^k-\mu_j^k\|\leq\| p_j^k-\lambda_j^k\|+\|\lambda_j^k-\mu_j^k\|=\zeta^k\|c\|+\|\lambda_j^k-\mu_j^k\|$,  we have $\|\theta_j^{k+1}-\mu_j^k\|\leq\zeta^k\|c\|+\|\lambda_j^k-\mu_j^k\|$. By taking limits on both sides, we obtain $\lim_{k\rightarrow\infty}\|\theta_j^{k+1}-\mu_j^k\|=0$.
 \end{pf}
The second part is a stochastically ``decreasing'' result, whose proof is similar to that of \citet[Lemma 4]{you2016scenario}, and is omitted here.
\begin{lem1}\label{lemma:stodec} (Stochastically decreasing). Let $\mathcal{F}^k$ be a $\sigma$-field generated by the random variable $\{\omega_j^k,j\in\mathcal{V}\}$ up to time $k$. Under Assumption \ref{assum:solvable} and \ref{assum:random}, it holds almost surely that for $\forall j\in\mathcal{V}$ and sufficiently large number $k$,
\begin{align}
\mathbb{E}&\left[\|\theta_j^{k+1}-\theta^*\|^2\right|\mathcal{F}^k]\notag\\
\leq&\ph (1+A_1{(\zeta^k)}^2)\|\lambda_j^k-\theta^*\|^2-2\zeta^kc^T(\mu_j^k-\theta^*)\notag\\
\label{equ:l5res}
&\ph -A_2\|\lambda_j^k-\mu_j^k\|^2+A_3{(\zeta^k)}^2.
\end{align}
where $\lambda_j^k$, $\mu_j^k$ are given in \myref{equ:l4def}, $\theta^*\in\Theta^*$, and $A_1, A_2, A_3$ are positive constants.
\end{lem1}
Finally, we can show that the consensus value is a weighted average of the state vector in each node, of which the weighted vector is the Perron vector of $A$.
\begin{lem1}\label{lemma:consensus} \citep{you2016scenario}. Consider the following sequence
\begin{equation}
\label{equ:l6def}
\theta_j^{k+1}=\sum_{i=1}^na_{ji}\theta_i^k+\epsilon_j^k,\quad\forall j\in\mathcal{V}.
\end{equation}
Suppose that $\mathcal{G}$ is strongly connected, let $\bartheta^k=\sum_{i=1}^n\pi_i\theta_i^k$, where $\pi_i$ is an element of $\pi$ given by \myref{equ:l1res1}. If $\lim_{k\rightarrow\infty}\|\epsilon_j^k\|=0$, it holds that
\begin{equation}
\label{equ:l6res}
\lim_{k\rightarrow\infty}\|\theta_j^k-\bartheta^k\|=0,\quad\forall j\in\mathcal{V}.
\end{equation}
\end{lem1}
The proof also relies crucially on the well-known super-martingale convergence Theorem, which is due to \citet{robbins1985convergence}, see also \citet[Proposition A.4.5]{bertsekas2015convex}. This result is now restated for completeness.
\begin{thm1}\label{thm:superm} (Super-martingale Convergence). Let $\{v_k\}$, $\{u_k\}$, $\{a_k\}$ and $\{b_k\}$ be sequences of nonnegative random variables such that
\begin{equation}
\mathbb{E}\left[v_{k+1}|\mathcal{F}_k\right]\leq (1+a_k)v_k-u_k+b_k
\end{equation}
where $\mathcal{F}_k$ denotes the collection $v_0,\dots,v_k$, $u_0,\dots,u_k$, $a_0,\dots,a_k$, $b_0,\dots,b_k$. Let $\sum_{k=0}^\infty a_k<\infty$ and $\sum_{k=0}^\infty b_k<\infty$ almost surely. Then, we have $\lim_{k\rightarrow\infty}v_k=v$ for a random variable $v\geq0$ and $\sum_{k=0}^\infty u_k<\infty$ almost surely.
\end{thm1}
Now, we can summarize the previous discussions.
\begin{prop1}\label{prop:1} (Convergent Results). Let $\barlambda^k=\sum_{j=1}^n\pi_j\mu_j^k$, and $\barmu^k=\sum_{j=1}^n\pi_j\mu_j^k$, where $\lambda_j^k$, $\mu_j^k$ are defined in \myref{equ:l4def} and $\pi$ is given in \myref{equ:l1res1}. Then, for any $\theta^*\in\Theta^*$, the following statements hold almost surely.\\
(a) $\{\sum_{j=1}^n\pi_j\|\theta_j^k-\theta^*\|^2\}$ converges.\\
(b) $\lim\inf_{k\rightarrow\infty}c^T\barmu^k=c^T\theta^*$.\\
(c) $\lim_{k\rightarrow\infty}\|\mu_j^k-\lambda_j^k\|=0$.\\
(d) $\lim_{k\rightarrow\infty}\|\mu_j^k-\theta_j^{k+1}\|=\lim_{k\rightarrow\infty}\|\lambda_j^k-\theta_j^{k+1}\|=0$.\\
(e) $\lim_{k\rightarrow\infty}\|\barmu^k-\bartheta^{k+1}\|=\lim_{k\rightarrow\infty}\|\barlambda^k-\bartheta^{k+1}\|=0$.
\end{prop1}
\begin{pf} By the convexity of $\|\cdot\|^2$ and the row stochasticity of $A$, i.e, $\sum_{i=1}^na_{ji}=1$, it follows that
\begin{equation*}
\|\lambda_j^k-\theta^*\|^2\leq\sum_{i=1}^na_{ji}\|\theta_i^k-\theta^*\|^2.
\end{equation*}
Jointly with \myref{equ:l5res}, we obtain that  for sufficiently large $k$,
\begin{align}
\mathbb{E}&\left[\|\theta_j^{k+1}-\theta^*\|^2\mathcal{F}^k\right]\notag\\ \leq&\ph(1+A_1{(\zeta^k)}^2)\sum_{i=1}^na_{ji}\|\theta_i^k-\theta^*\|^2-2\zeta^kc^T(\mu_j^k-\theta^*)\notag\\
\label{equ:l7pr1}
&\ph -A_2\|\lambda_j^k-\mu_j^k\|^2+A_3{(\zeta^k)}^2.
\end{align}
We multiply both sides of \myref{equ:l7pr1} with $\pi_j$ and sum over $j$, together with \myref{equ:l1res1} and the definition of $\barmu^k$, we obtain
\begin{align}
\mathbb{E}&\left[\sum_{j=1}^n\pi_j\|\theta_j^{k+1}-\theta^*\|^2|\mathcal{F}^k\right]\notag\\
\leq&\ph (1+A_1{(\zeta^k)}^2)\sum_{j=1}^n\sum_{i=1}^n\pi_ja_{ji}\|\theta_i^k-\theta^*\|^2-2\zeta^kc^T(\barmu^k-\theta^*)\notag\\
&\ph -A_2\sum_{j=1}^n\pi_j\|\lambda_j^k-\mu_j^k\|^2+A_3{(\zeta^k)}^2\notag\\
=&\ph (1+A_1{(\zeta^k)}^2)\sum_{j=1}^n\pi_j\|\theta_j^k-\theta^*\|^2-2\zeta^kc^T(\barmu^k-\theta^*).\notag\\
&\ph -A_2\sum_{j=1}^n\pi_j\|\lambda_j^k-\mu_j^k\|^2+A_3{(\zeta^k)}^2.\notag
\end{align}
It follows from \myref{equ:stepdef} that $A_1{(\zeta^k)}^2\geq0$, $A_3{(\zeta^k)}^2\geq0$, $\sum_{k=0}^\infty A_1{(\zeta^k)}^2<\infty$ and $\sum_{k=0}^\infty A_3{(\zeta^k)}^2<\infty$ hold. Notice the convexity of $\Theta_0$ and  $\mu_j^k\in\Theta_0$, it is clear that $\barmu^k\in\Theta_0$. In view of the fact that $\theta^*$ is one optimal solution in $\Theta_0$, it holds that $c^T\barmu^k-c^T\theta^*\geq 0$. Thus, all the conditions in Theorem \ref{thm:superm} are satisfied. It holds almost surely that $\{\sum_{j=1}^n\pi_j\|\theta_j^k-\theta^*\|^2\}$ is convergent for any $j\in\mathcal{V}$ and $\theta^*\in\Theta^*$, hence (a) is proved. Moreover, it follows from Theorem \ref{thm:superm} that
\begin{equation}
\label{equ:l7rb}
\sum_{k=0}^\infty\zeta^kc^T(\barmu^k-\theta^*)<\infty
\end{equation}
and
\begin{equation}
\label{equ:l7rc}
\sum_{k=0}^\infty\sum_{j=1}^n\pi_j\|\lambda_j^k-\mu_j^k\|^2<\infty.
\end{equation}
It is clear that \myref{equ:l7rb} directly implies (b) under the condition $c^T\barmu^k-c^T\theta^*\geq0$. Together with the fact that $\pi_i>0$ from Lemma \ref{lemma:perron}, it follows from \myref{equ:l7rc} that $\lim_{k\rightarrow\infty}\|\lambda_j^k-\mu_j^k\|^2=0$, thus (c) is proved. Combining the result of (c) with Lemma \myref{lemma:feasi}, it is clear that (d) holds as well. As for (e), it is the direct inference of (d) by using triangle inequality, i.e., $\| \barlambda^k-\bartheta^{k+1}\|\leq\sum_{j=1}^n\pi_j\|\lambda_j^k-\theta_j^{k+1}\|$.
\end{pf}
Combining the above results, we are in a position to formally prove Theorem \ref{thm:result}.

{\bf Proof of Theorem \ref{thm:result}}. Notice that $\lambda_j^k=\sum_{i=1}^na_{ji}\theta_j^k$, it follows from Proposition 1(d) that $\lim_{k\rightarrow\infty}\|\theta_j^{k+1}-\sum_{i=1}^na_{ji}\theta_j^k\|=0$. Then it holds almost surely from Lemma \ref{lemma:consensus} that $\lim_{k\rightarrow\infty}\|\theta_j^k-\bartheta^k\|=0$. Jointly with the fact Proposition 1(a) that $\{\sum_{j=1}^n\pi_j\|\theta_j^k-\theta^*\|^2\}$ converges for any $\theta^*\in\Theta^*$, we obtain that $\{\|\bartheta^k-\theta^*\|\}$ converges almost surely for any $\theta^*\in\Theta^*$. Then it follows from Proposition 1(e) that $\{\|\barmu^k-\theta^*\|\}$ converges as well. By Proposition 1(b), it implies that there exists a subsequence of $\{\barmu^k\}$ that converges almost surely to some point in the optimal set $\Theta^*$, which is denoted as $\theta_{opt}^*$. Due to the convergence of $\{\|\barmu^k-\theta_{opt}^*\|\}$, it follows that $\lim_{k\rightarrow\infty}\barmu^k=\theta_{opt}^*$. Finally, we note that $\|\theta_j^{k+1}-\theta_{opt}^*\|\leq\|\theta_j^{k+1}-\bartheta^{k+1}\|+\|\bartheta^{k+1}-\barmu^k\|+\|\barmu^k-\theta_{opt}^*\|$, which converges almost surely to zero as $k\rightarrow\infty$. Therefore, there exists $\theta_{opt}^*\in\Theta^*$ such that $\lim_{k\rightarrow\infty}\theta_j^k=\theta_{opt}^*$ for all $j\in\mathcal{V}$ with probability one. Thus, Theorem \ref{thm:result} is proved.
\section{Illustrative Examples}
\label{experiment}
We consider the facility location problem, which is one of the classical problems in operations research. Traditional facility location is a centralized problem, while in this paper, we propose a distributed formulation of the problem.
\begin{align}
\min_{x\in\mathbb{R}^2}&\sum_{i=1}^nw_i\|x-q_i\|\notag\\
\text{s.t.  } &\|x-p_i^{(1)}\|\leq l_i^{(1)},\notag\\
\label{equ:apply}
&\|x-p_i^{(2)}\|\leq l_i^{(2)},\quad i=1,...,n
\end{align}
The local constraints in \myref{equ:apply} represent local resource limitations in each node, and the objective function describes the sum of cost when the facility is settled.

\begin{figure}[htp]
\centering
\includegraphics[width=4cm]{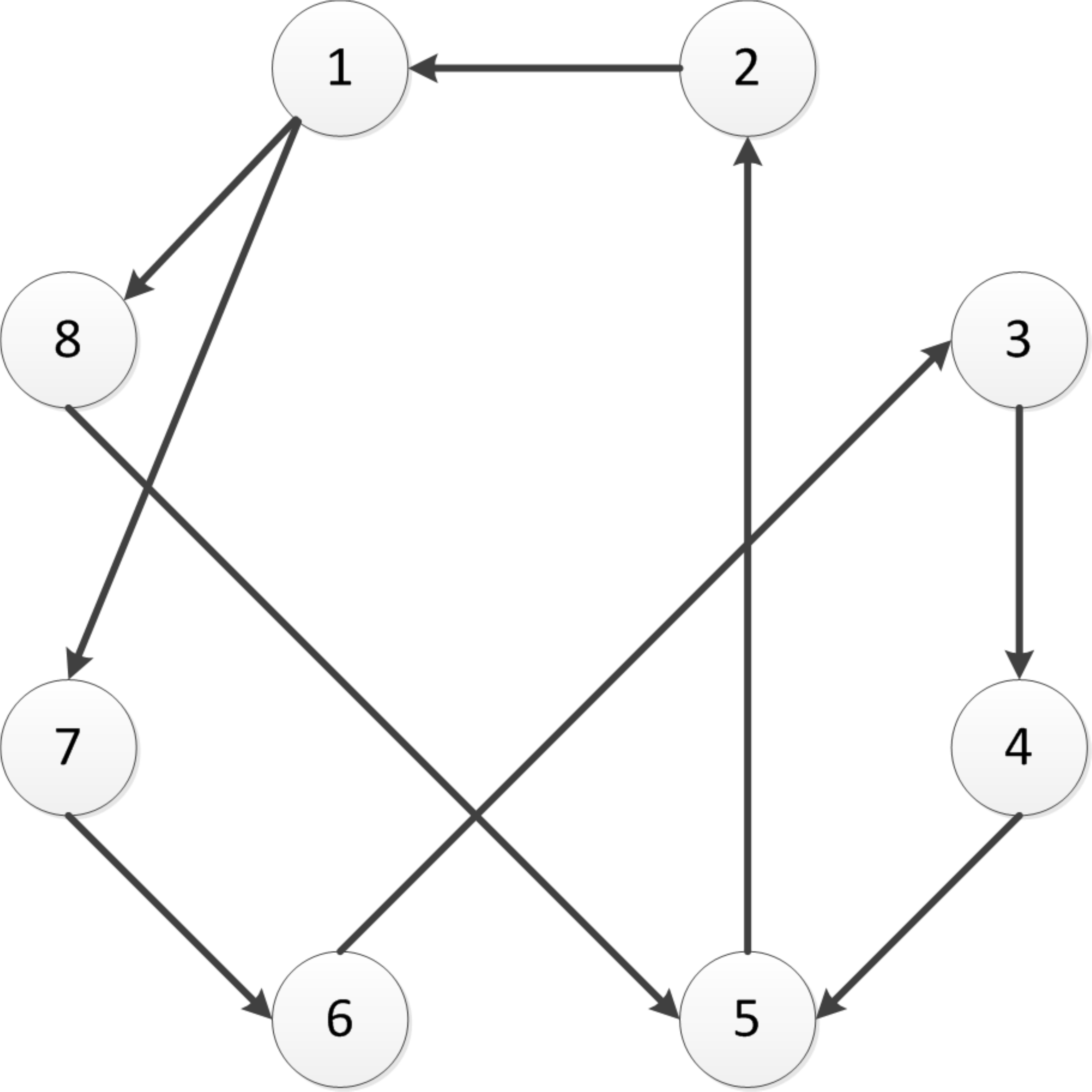}
\caption{A strongly connected but unbalanced digraph.}
\label{fig:fixedgraph}
\end{figure}

\begin{figure}[htp]
\centering
\includegraphics[width=8.4cm]{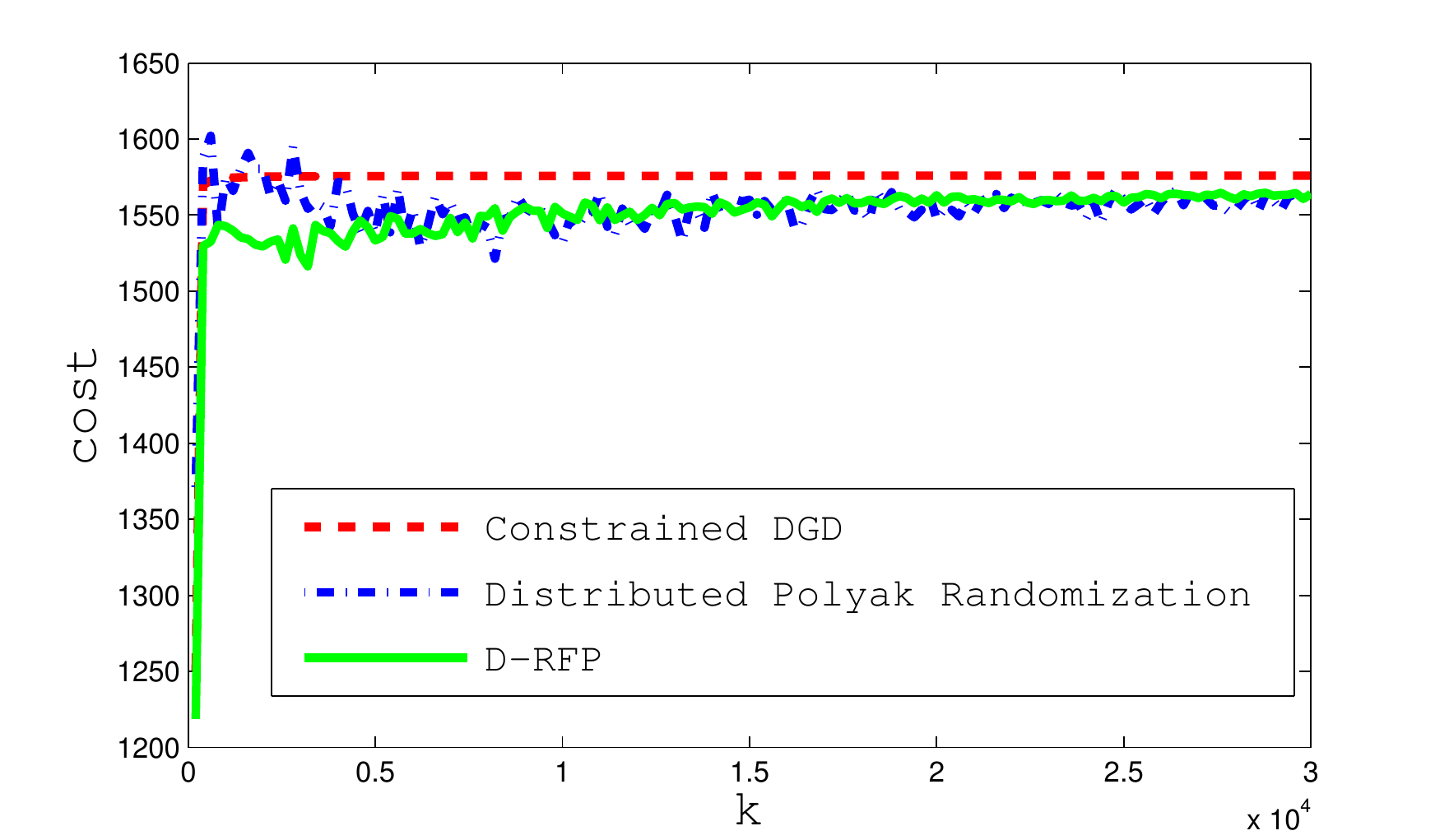}    
\caption{Comparisons of constrained DGD, Distributed Polyak randomization, and D-RFP.}
\label{fig:comparison}
\end{figure}

\begin{figure}[htp]
\centering
\includegraphics[width=4cm]{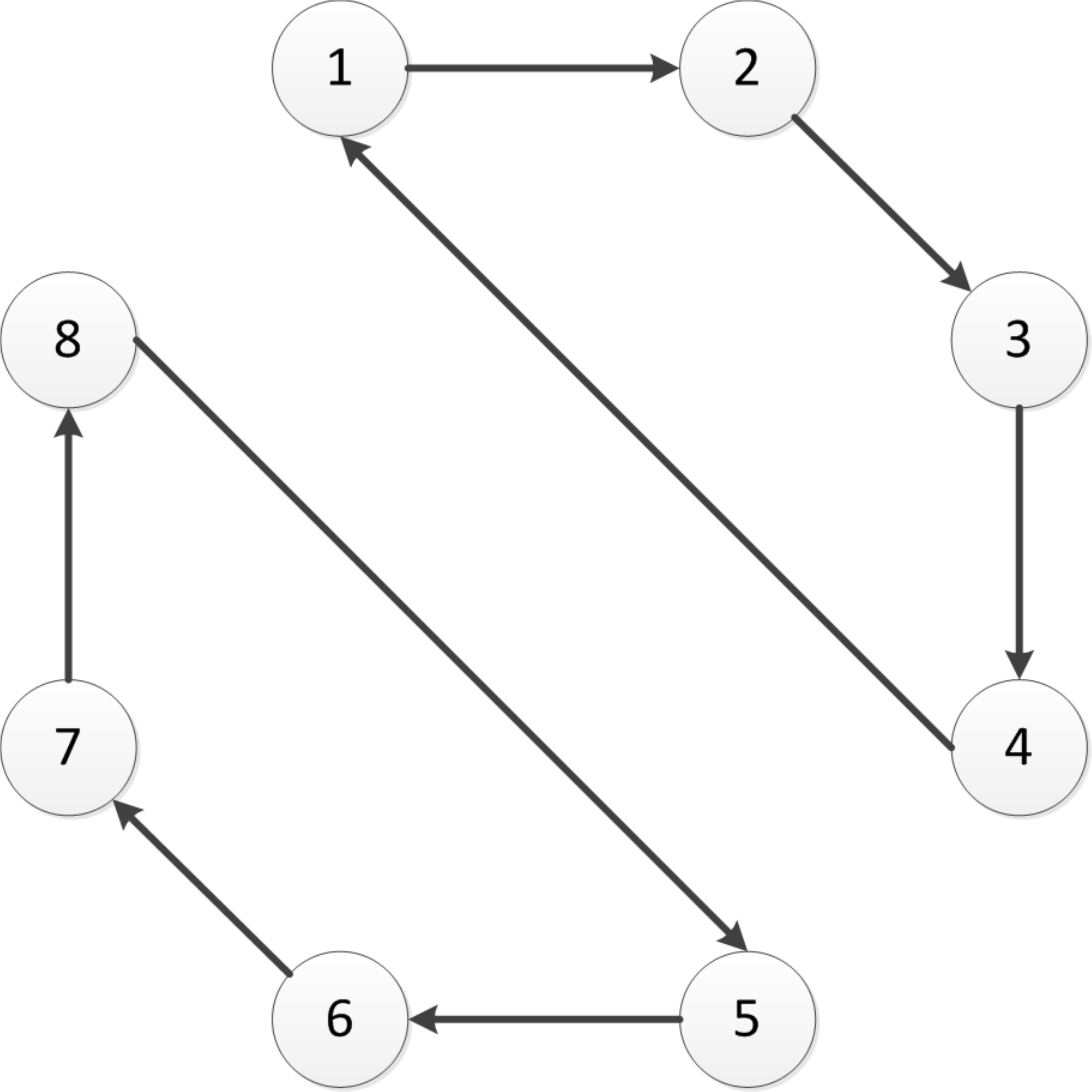}
\quad
\includegraphics[width=4cm]{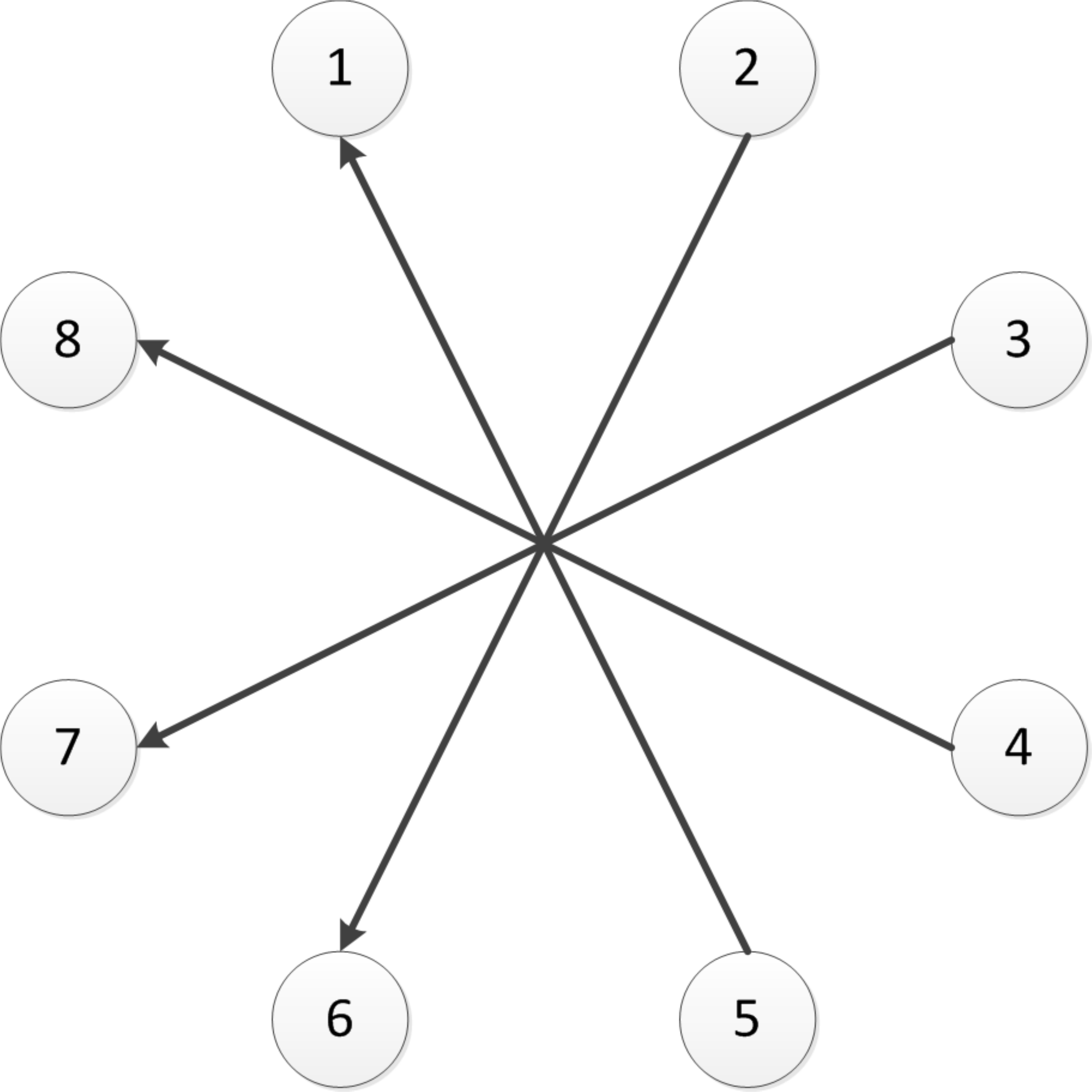}
\caption{Two jointly strongly connected digraphs, but individually not strongly connected.}
\label{fig:switchinggraph}
\end{figure}

\begin{figure}[htp]
\centering
\includegraphics[width=8.4cm]{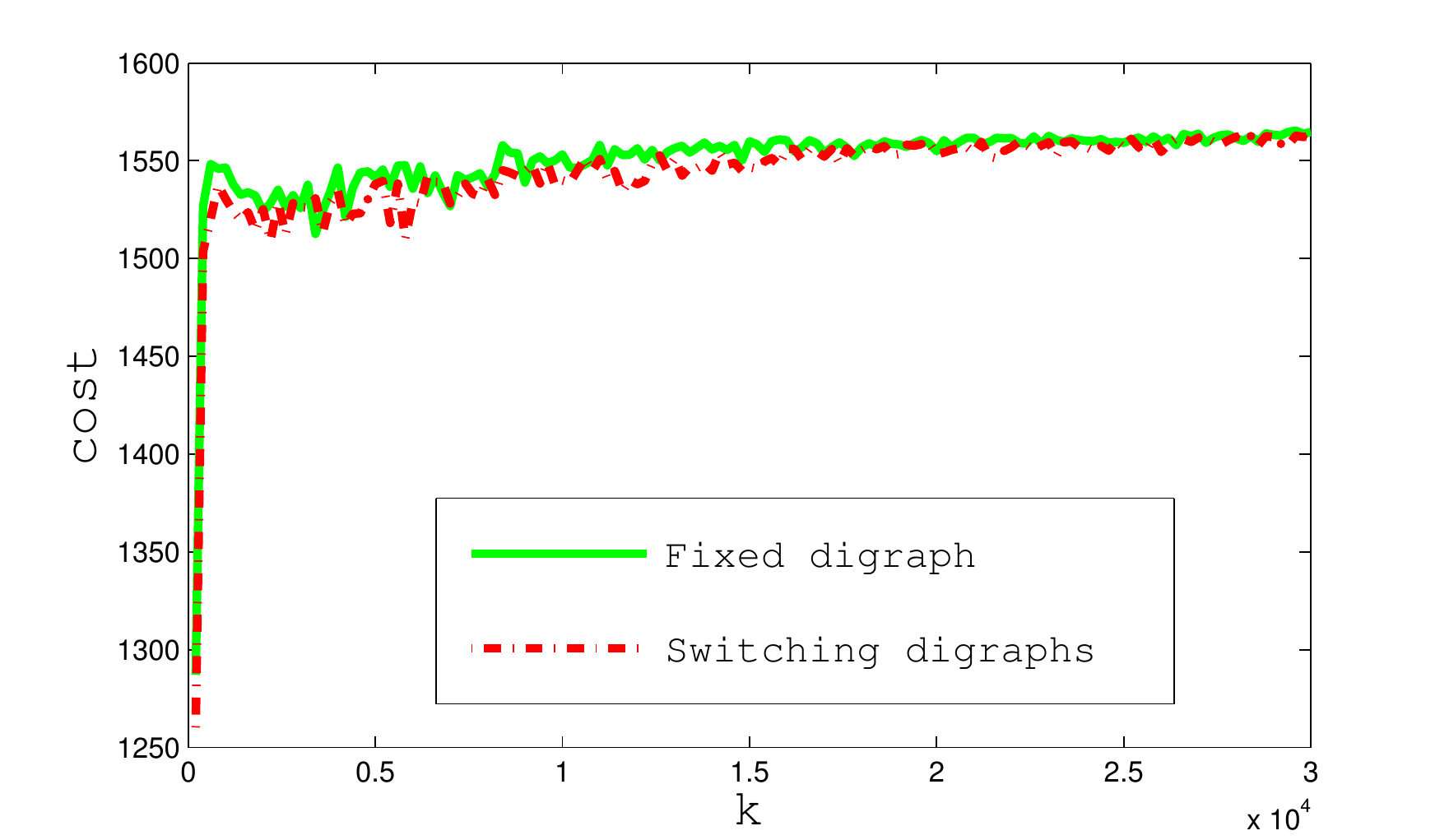}
\caption{The D-RFP over time-varying digraphs and its comparison to the case under a fixed digraph.}
\label{fig:time-varying}
\end{figure}

We first compare several algorithms over a strongly-connected directed graph (omit self-loops) in Fig. \ref{fig:fixedgraph}.
The associated weighting matrix is only row-stochastic. In this experiment, three algorithms are performed under the same stepsizes, e.g.,  $\beta=1,\ph\zeta^k=\frac{1}{k}$. Comparisons of D-RFP, distributed Polyak randomization \citep{you2016scenario} and the constrained extension of DGD \citep{nedic2010constrained} are shown in Fig. \ref{fig:comparison}.  We can clearly observe that the minimal cost calculated by the constrained DGD is greater than the other two algorithms, which implies that the constrained DGD does not converge to an optimal solution. This is consistent with the observation in Section \ref{subsec_review}. The result in Fig. \ref{fig:comparison}  indicates that the D-RFP converges faster, and is with less fluctuations than the distributed version of the Polyak randomization algorithm. This is mainly due to the use of an additional fixed ``projection' in D-RFP.
We also apply the D-RFP to time-varying unbalanced digraphs. For time-varying graphs, a common assumption is that the graph sequence $\{\mathcal{G}(t)\}$ is uniformly strongly connected \citep{nedic2015distributed}, i.e., there exists a constant $L$ such that $\mathcal{G}(t)\cap...\cap\mathcal{G}(t+L)$ is strongly connected for any $t>0$. In this experiment, the time-varying graphs are given in  Fig. \ref{fig:switchinggraph}, where any graph is not strongly connected but their joint graph is strongly connected. We assume that $\mathcal{G}(t)$ is the left graph at odd time, and otherwise, is the right one. The simulation result in Fig. \ref{fig:time-varying}  confirms that our algorithm is also applicable to time-varying digraphs. The proof will be included in the journal version.

\section{CONCLUSIONS}
\label{conclusion}
In this work, we developed a random-fixed projected algorithm  to collaboratively solve distributed constrained optimizations over unbalanced digraphs. The proposed algorithm has a simple structure. The simulation indicates that the proposed algorithm is applicable to time-varying digraphs, of which the strict proof will be given in the journal version. The drawback of our algorithm is that the number of the augmented variables depends on the scale of topology, which is an open question. Future work will focus on reducing the number of augmented variables and accelerating the convergence speed.
\bibliography{ifacconf}             







\end{document}